\documentclass{mem}
\usepackage{natbib}\usepackage{txfonts}\usepackage{balance}
\usepackage{graphicx}
\usepackage[a4paper,breaklinks,dvipdfm]{hyperref}
\idline{75}{282}
\begin{document}

\title{My chemistry with Francesco}

\subtitle{}

\author{D. \,Galli}
 
\institute{INAF -- Osservatorio Astrofisico di Arcetri, Largo E. Fermi 5,
I-50125 Firenze, Italy \\
\email{galli@arcetri.astro.it}}

\authorrunning{Galli}

\titlerunning{My chemistry with Francesco}

\abstract{Through all his career, Francesco maintained a keen 
interest in primordial star formation and 
the chemistry of the early Universe. It was therefore quite natural for me, 
his former student and his officemate for more than 12 years, 
to be also involved in these studies. In this contribution I summarize the chemistry that 
Francesco and I developed, pointing out the main findings and false steps 
of our lifelong collaboration.

}

\maketitle{}

\section{Introduction}

\begin{figure}[t!]
\resizebox{\hsize}{!}{\includegraphics[angle=-90]{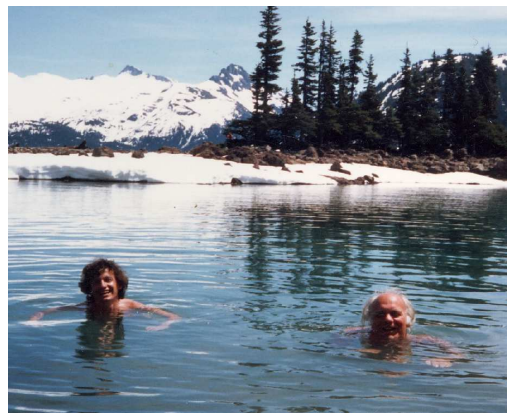}}
\caption{\footnotesize Francesco Palla and Franco Pacini swim in the Garibaldi Lake,
British Columbia (July 1987).
}
\label{fpfp}
\end{figure}

In the summer of 1987, Ralph Pudritz and Michel Fich organized a NATO School 
on ``Galactic and Extragalactic Star Formation'' in 
Whistler, a resort  town in beautiful British Columbia, where Francesco was invited to
give a review lecture on ``Primordial Star Formation". Among the participants 
to the School were three PhD student of the University of Florence 
(Riccardo Cesaroni, Paolo Lenzuni and myself), and the somewhat more mature 
Director of the Arcetri Observatory (Franco Pacini).
For the three of us, the youngest members of this scientific expedition, the 1987 NATO School was 
the first experience of an international astronomical conference, and, in my case,
my first exciting contact with the great American continent. A picture taken during the School
(Fig.~\ref{fpfp}) shows a relaxing moment during one of our excursions, when Francesco 
and Franco Pacini engaged in a resistance contest in the freezing-cold waters 
of the Garibaldi Lake (for the records, Francesco lost).
Francesco's lecture, published one year later (Palla~1988), stimulated my interest 
in the chemistry of the early Universe, and our discussions during long walks in the 
Canadian woods marked the beginning of a lifelong collaboration with him
on this subject. 

It is instructive to look in retrospect at the topics that Francesco addressed
in his 1987 lecture, to assess the progress (or the lack of it) and the 
expected (or unexpected) developments that have occurred in this field in the 
intervening 30 years. Francesco's lecture covered four broad subjects:
(1) the  search for zero-metal stars, (2) the chemistry of the early Universe,
(3) the evolution of collapsing gas clouds, and (4) the formation of primordial stars.
These topics are still in the forefront of current research, as also witnessed
by these Proceedings.
In his lecture, Francesco gave also a summary of his own
work done in collaboration with Ed Salpeter and Steve Stahler
that had been published only a few years before: the famous Palla-Salpeter-Stahler
(and permutations) ``trilogy'' of 1983--1986 (Palla et al.~1983;
Stahler et al.~1986a,b)

Our collaboration on primordial chemistry developed from these premises.
The first paper of the trilogy (which is Francesco's most cited paper) pointed 
out the importance of 3-body reactions, 
$3{\rm H}\rightarrow {\rm H}_2+{\rm H}$ and 
$2{\rm H}+{\rm H}_2\rightarrow {\rm H}_2+{\rm H}_2$
for the gas-phase formation of molecular hydrogen. The conversion of H into
H$_2$ was followed with the help 
of a very simple model for the collapse of a zero-metal cloud.
Francesco was aware of the limitations of the collapse model 
and the chemical network 
adopted, and wanted to improve both. Even though his main interest was on 
the former aspect (collapse and star formation), we resolved to work 
first on the latter (the chemistry), and continued to do so until 2013. The motivation 
for proceeding in this order was
the poor state of primordial chemistry at the time, with many reaction channels 
poorly identified, reactions rates badly guessed and often largely discrepant 
from author to author, and the relevant data scattered
in publications not easily accessible (it was before the ADS).
Especially annoying was the uncertainty on the H$_2$ cooling rate for 
collisions with H atoms. In the 1980's, two 
independent calculation of this fundamental ingredient for
any recipe of primordial star formation were available: 
one by Lepp \& Shull~(1983), and one by Hollenbach \& McKee (1979, revised
in 1989), based on independent sets of collisional rate coefficients.
The problem was that the two cooling rates in the low-density 
regime differed by as much as one order of magnitude or more 
below $T\approx 100$~K, a temperature range 
relevant for collapse calculations. As a result, 
the estimate of the minimum mass needed to collapse at redshift $z\approx 30$--40
was uncertain by a factor of $\sim 30$ (see e.g. Fig.~7 of Palla~1999).
Clearly, 
before attacking our planned collapse calculation,
a systematic and critical reanalysis of the microphysics was in order.
The collapse calculation we originally planned was eventually done, but not by us. The work
by Kazuyuki Omukai and coworkers is in my opinion the ideal fulfilment of 
our program (Omukai \& Nishi~1998, Omukai~2000).
Later on, Francesco would ``exploit'' K.~Omukai's
capabilities to further develop his trilogy papers: 
first, to explore the formation of massive stars by the enhanced 
accretion rate expected in primordial conditions (Omukai \& Palla~2001);
and, second, to extend his 1986 calculation of the mass-radius relation for protostars
to the case of a zero-metallicity cloud (Omukai \& Palla~2003). 

Turning back to our chemistry, our critical selection and analysis of  gas-phase
reaction rates for a mixture of H, He, D, Li and their
products was eventually completed and published in 1998, together with our ``new and improved'' cooling
rates for the main molecular species (Galli \& Palla~1998, hereafter GP98).
Our recommended reaction and cooling rates were
widely adopted by researchers in the field 
(GP98 is my most cited paper), showing 
that the humble work of reordering and systematizing can be useful too. 
Ten years later, the H-H$_2$ cooling rate by GP98 was superseded by a newer calculation by Glover \& Abel~(2008),
based on updated collisional rate coefficients. The last word on the subject is represented in my opinion 
by the extensive set of theoretical calculations performed by Fran\c cois Lique and coworkers (Lique et al.~2012; 
Lique~2015) of H-H$_2$ collisional cross sections and rate coefficients down to temperatures of 10~K, whose impact 
on the H-H$_2$ cooling rate still needs to be fully assessed. 

In 2013 Francesco and I were invited to write an Annual Reviews paper on primordial chemistry.
For us this was a good opportunity to summarize the latest developments in a 
historical context (Galli \& Palla~2013). We realized that in the intervening years many of the uncertainties that affected 
the results of GP98 had been largely overcome: first of all, cosmology had entered the ``precision era'', 
and the uncertainty on the chemistry of the Dark Ages was no longer dominated 
by uncertainties in the values of the cosmological parameters (all accurately determined 
by Boomerang, Planck and WMAP), but by residual uncertainties in the chemical 
reaction rates. These, fortunately, were found to
be very small at least for the key reactions, with the possible exception of the 3-body 
reactions mentioned earlier, and others of minor relevance. The 2013 review was the last
paper I had the privilege and the pleasure of writing sitting side-by-side with Francesco.
It was, as always, an enjoyable personal experience: Francesco was dead 
serious about science, careful and rigorous in his work, but at the same time light-hearted, 
playful and full of irony with his coworkers.

Francesco and I were also curious
about the possible observational signatures left by primordial molecules in the form of both spatial
fluctuations and spectral distortions in the cosmic microwave background (CMB). This 
idea was first advanced  by Viktor Dubrovich in the 1970's (Dubrovich~1977),
endorsed by Zel'dovich~(1978), and 
later developed in depth by Prof.~Francesco Melchiorri and his group at the University
La~Sapienza in Rome, where Francesco had studied (Maoli et al.~1994, 1996). 
The attractive feature of this proposal 
was that a hypothetical absorbing layer of molecules extending
over a significant range of redshifts could erase or attenuate primary anisotropies produced at higher
redshift, reconciling cosmological models of structure formation with the small magnitude and large angular
scale of the spatial anisotropies known at the time. The attention 
concentrated on the LiH molecule, due to its high dipole moment, and its (expected) 
efficient formation by radiative association. Unfortunately, 
the latter process was found to occur at a very slow rate
(Dalgarno et al.~1996; Gianturco \& Gori Giorgi~1997), too slow for LiH to play any
significant role in the early Universe. It was however exciting for Francesco and myself
to get involved in the laboratory frequency measurements of 14 rotational transitions of LiH
in the fundamental and in the first two excited vibrational states (Bellini et al.~1994),  
and collaborate with Prof.~Francesco Gianturco of the University of Rome and 
his student Stefano Bovino on accurate 
determinations of reaction rates relevant for the gas-phase chemistry of Li and He 
(Bovino et al.~2011a,b; 2012). Then,
in collaboration with a young brilliant post-doc, Dominik Schleicher, who
visited the Arcetri Observatory during 2008,
we computed the distortions imprinted in the CMB by the chemistry
taking place during the Dark Ages (Schleicher et al.~2008, 2009). The changes
in the CMB intensity and power spectrum were found to be too small to be detected by Planck,
but within reach of planned future CMB satellite like PIXIES and PRISM.  The most promising
species from this point of view were identified in H$^-$ (as also suggested by Black~2006) 
and HeH$^+$. A few years later Francesco and I met another bright freshly graduated young
researcher, Carla Maria Coppola of the University of Bari, who patiently and skilfully educated both of 
us through the intricacies of state-resolved chemistry, in particular of H$_2$ and H$_2^+$,
and the subtle effects of non-thermal (recombination) photons 
(Coppola et al. 2011, 2012, 2013, 2016, 2017). Our work also benefited from many discussions
with Raffaella Schneider, first met by Francesco and myself at a cosmology meeting in Frascati 
in 1996, where a just graduated Raffaella
delivered her talk with a mix of intelligence and grace that 
has remained unchanged over the years.

My chemistry with Francesco was not limited to primordial molecules: we were both very curious
about primordial isotopes as well, in particular D and $^3$He (Galli et al.~1995). 
The latter isotope was especially intriguing, and a lot 
of effort was devoted to understand why, if $^3$He is produced by low-mass stars
as predicted by stellar nucleosynthesis models, its
present-day abundance measured in H{\sc ii} regions and the local ISM, as well as in 
gas-rich meteorites, is about one order of magnitude lower than it should be. It was 
an interesting problem, since this discrepancy, the so-called ``$^3$He problem'' hampered
the use of this isotope as a cosmic baryometer along with D, $^4$He and $^7$Li (the latter
being problematic as well, see Paolo Molaro's contribution in these Proceedings). 
We found an elegant nuclear physics solution to this problem (Galli et al.~1994), but the 
ugly truth was eventually discovered by Charbonnel (1995) and Hogan (1995) 
in the form of a non-standard mixing mechanism acting in low-mass stars during the red-giant branch 
evolution or later, leading to a substantial (or complete) destruction of all their freshly
 produced $^3$He. However, we were in some sense vindicated by the observational evidence that at least 
 {\em some} stars do not appear to suffer extra-mixing during all their lives (Galli et al.~1997). 
 
This was my chemistry with Francesco: curiosity-driven research,
and the pure pleasure of sharing knowledge with others. 
My chemistry partner has truly enriched my life.

\bibliographystyle{aa}

\end{document}